\documentclass[prl,twocolumn,aps,superscriptaddress,showpacs,floatfix]{revtex4}
\usepackage{amssymb}

\usepackage{eurosym}
\usepackage{graphicx}
\usepackage{color}

\begin{document}

\title{Perfect memristor: non-volatility, switching and negative
differential resistance}
\author{Sergey E. Savel'ev}
\affiliation{Department of Physics, Loughborough University, Loughborough LE11 3TU,
United Kingdom}
\author{Fabio Marchesoni}
\affiliation{Dipartimento di Fisica, Universit\'a di Camerino, I-62032 Camerino, Italy}
\author{Alexander M. Bratkovsky}
\affiliation{Hewlett-Packard Laboratories, 1501 Page Mill Road, California 94304}
\affiliation{Department of Physics, University of California, Davis, One Shields Avenue,
Davis, California 95616}
\affiliation{Kapitza Institute for Physical Problems, Russian Academy of Sciences,
Kosygina 2, 119334 Moscow, Russia}

\begin{abstract}
We propose a simple model of a nanoswitch as a memory resistor. The
resistance of the nanoswitch is determined by electron tunneling through a
nanoparticle diffusing around one or more potential minima located between
the electrodes in the presence of Joule's heat dissipation. In the case of a
single potential minimum, we observe hysteresis of the resistance at finite
applied currents and a negative differential resistance. For two (or more)
minima the switching mechanism is non-volatile, meaning that the memristor
can switch to a resistive state of choice and stay there. Moreover, the
noise spectra of the switch exhibit $1/f^2\rightarrow 1/f$ crossover, in
agreement with recent experimental results.
\end{abstract}

\maketitle

Fast progress in flash memory suggests that standard magnetic hard drives
will soon be replaced by faster and much more reliable solid state drives 
\cite{silicon}. However, solid state drives are still very expensive and
their capacitance is very limited when compared with magnetic drives. To
overcome these limitations, a new generation of flash-like memory has to be
developed \cite{book}.

\begin{figure}[tbp]
\includegraphics[width=9cm]{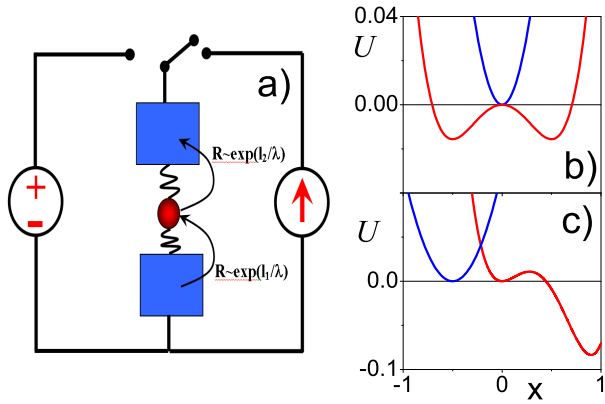}
\caption{Schematic diagram of the proposed memristor 
driven by either a current or voltage source: (a) A nanoparticle
(solid red circle) can fluctuate between two electrodes (blue squares), due
to thermal noise. This results in a fluctuating resistance of electrons
tunneling through the system (see text). (b),(c) The nanoparticle is subject
to either a single well or a double well potential, mirror symmetric in (b)
and asymmetric in (c).}
\label{F1}
\end{figure}

A most promising candidate for very fast and high density memory are
nanomemristors (memory-resistors) \cite{stan}. In particular, resistive
switching was observed in binary oxides, like TiO$_{2}$, NiO and perovskite
oxide memristors. Memristive switches would provide a viable solution for
dense memory, logic, and neurocomputing, if they can be demonstrated to
advantageously scale with size, power, and driving voltage, and perform
without failure and fatigue in a repeatable and uniform manner. Some of
these issues are not well understood although the hysteretic behavior of
such materials, especially thin films of transition metal oxides like Ta$_{2}
$O$_{5}$, Nb$_{2}$O$_{5}$, TiO$_{2}$, NiO, Cu$_{2}$O in
metal-insulator-metal vertical devices has been known for decades \cite{1}.
In recent years, the interest in various oxide-based systems switchable by
electric pulses has grown dramatically \cite{2,3,4} leading to important
breakthroughs, despite a still poor understanding of the switching
mechanisms and, in particular, of the joint heat-electron-ion transport at
the nanoscale.

Molecular dynamics simulations \cite{5,6} of interacting oxygen vacancies
have demonstrated the formation of vacancy filaments and clusters
characterized by a certain degree of orientational order. This picture is in
qualitative agreement with recent measurements \cite{7}, which hint at the
existence of localized conducting channels. Moreover, the formation of
nanoswitches (buried atomic junctions) between large clusters of vacancies
would be consistent with the observed quantum conductance and the unusual
resistance-noise spectrum in TaO$_{x}$ memristors \cite{6}. The question
then arises to what extent nanoswitches formed in oxide memristors are
responsible for resistive switching and how to effectively control switching
in such nanojunctions.

In this Letter, we consider the simplest possible model of joint
electron-ion-heat dynamics for a nanoswitch with electrons tunneling through
a bridging nanoparticle (either an ion or a vacancy) subject to thermal
fluctuations. Electric current through the switch produces Joule heating,
which results in local temperature relaxation, described by Newton's cooling
law. Surprisingly, we demonstrate that this system admits two distinct
resistive states even if the nanoparticle has only one mechanical attractor
(e.g., the particle moves in a potential with a single minimum). 
Moreover, if the particle is confined by a potential with, say, two minima,
these two resistive states can survive even at zero applied current (or
voltage), so that the junction could serve as a non-volatile resistive
switch. Furthermore, by applying current-voltage pulses such a memristor can
be made switch into the desired resistive state, almost deterministically,
although the particle is subject to random forces, only. Therefore, the resistance
of this switch can be cycled through its hysteresis loop in apparent analogy
with the elementary mechanism of magnetic memory devices. In this regard,
our model exhibits all key features of a resistive memory device.

\begin{figure}[tbp]
\includegraphics[width=8cm]{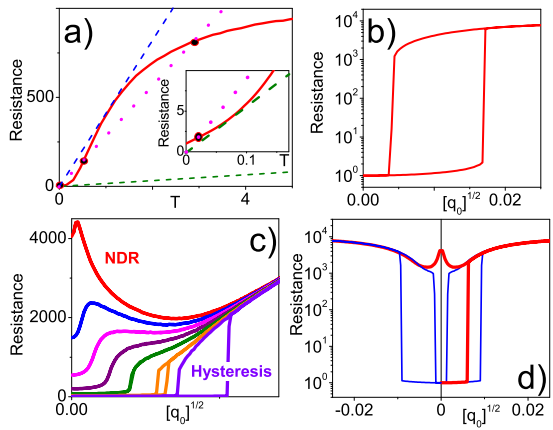}
\caption{Slow temperature relaxation: (a) Temperature dependence of the
memristor resistance, $R(T)=\langle R(t,T)\rangle $ simulated by using Eq. (\ref{1}) 
at fixed $T$. Black dots correspond to the stationary
points $\bar{r}(T_{s})=\kappa T_{s}/q_{0}$. Inset: blow-up of $R(T)$
at low $T$. (b) Simulations of coupled Eqs. (\ref{2}) and (\ref{1}) 
for a current driven memristor and a symmetric harmonic potential $U(x)$ 
[Fig. \ref{F1}(b)] with $\omega _{0}=2$. The
hysteretic loop of ${\bar{r}}$ vs. $[q_{0}]^{1/2}\propto I$ occurs only for
non-zero currents $I$. (c) Upon shifting the minimum of the parabolic
potential, $x_{0}$ off-center [Fig.\ref{F1}(c)], hysteresis is
gradually suppressed and negative differential resistance sets on; $x_{0}$
is increased from $0.3$ to $0.9$ by steps of $0.1$; (d) Resistance
hysteretic loops for a switch with asymmetric two-well potential, see text,
with $(\protect\alpha ,\protect\beta ,\protect\gamma )=(4,11.6,8)$ (blue
curve) and $(4,18.8,16)$ (red curves). Other simulation parameters are: time
integration step $\Delta t=0.005$, averaging time $t_{\max }=5\times 10^{4}$, 
$\protect\lambda =0.1$, $\protect\omega _{0}=2$, and $\protect\kappa =0.01$.}
\label{F2}
\end{figure}

\textit{Model.---\/} Let us consider a nanoparticle diffusing between two
electrodes subject to thermal noise (Fig. \ref{F1}a). An electric current
through the switch sets on when the electrons tunnel from one electrode to
the other via the nanoparticle. Here, we assume electron relaxation to be
much faster than heat diffusion and the nanoparticle dynamics, thus,
ignoring any shuttling effects \cite{shuttle}. Therefore, the instantaneous
resistance of the device, $R(x)$, is a function of the particle location, $x$, 
whereas its temperature, $T(x)$, results from the balance between Joule
heating, which depends on $R(x)$, and temperature relaxation. This
simplified electron-particle-heat dynamics can be formulated as 
\begin{eqnarray}
&&dx/dt+\partial U/\partial x=\sqrt{2T}\xi ,\ \ |x|<1,  \label{1} \\
&&dT/dt=q(R(x))-\kappa T,  \label{2}
\end{eqnarray}
where $x$ is the dimensionless coordinate of the nanoparticle, which is
confined by the binding potential $U(x)$ between the two electrodes at 
$x=\pm 1$ (normalization condition). The random force $\xi (t)$ represents a
Gaussian, zero-mean, white noise with $\langle \xi (0)\xi (t)\rangle =\delta
(t)$. In Eq. (\ref{2}) the time dependence of the temperature is described
by the standard Newton's cooling law with Joule heat source $Q(R)=Cq(R)$, $C$
and $\kappa $ being the junction heat capacitance and the temperature
relaxation rate, respectively. The tunneling resistance between the first
electrode and the particle and the particle and the second electrode are,
respectively, $R_{1}\propto \exp [(1-x)/\lambda ]$ and $R_{2}\propto \exp
[(x+1)/\lambda ]$, where $\lambda $ is the normalized \cite{norm} tunneling length. The
total switch resistance can be rewritten as $R=R_{1}+R_{2}=R_{0}r$, where 
$r=\cosh (x/\lambda )$ is a dimensionless resistance and $R_{0}$ is the
junction resistance when the particle is located at the midpoint between the
electrodes, $x=0$. If the particle is driven by an electrical current $I$,
Joule's law states that $Q=I^{2}R$, while in the case of voltage driven
switches, $Q=V^{2}/R$. Correspondingly, in compact notation, $q=q_{0}r^{n}$,
with $n=\pm 1$ and $q_{0}=I^{2}R_{0}/C$ or $q_{0}=V^{2}/R_{0}C$,
respectively, for current or voltage driven switches. The most interesting
working regimes occur when the temperature relaxation is either slow, 
$\kappa \ll 1,$ or fast, $\kappa \gg 1$.

\textit{Slow temperature relaxation.---\/} For a qualitative analysis in the
slow relaxation regime, the resistance can be averaged with respect to the
fast thermal fluctuations of the particle, 
$\langle R(x)\rangle ={\bar{R}}(T)=R_{0}{\bar{r}}(T)$, so that Eq. (\ref{2}) can be reduced to 
\begin{equation}
dT/dt=q_{0}{[\bar{r}(T)]}^{n}-\kappa T,  \label{3}
\end{equation}
with stationary points $T=T_{s}$ satisfying the identity 
$T_{s}=q_{0}{[\bar{r}(T_{s})]}^{n}/\kappa$.

We start our analysis by assuming that the nanoparticle moves in a simple
harmonic potential $U=\omega _{0}^{2}x^{2}/2$ [Fig. \ref{F1}(b)]. For a
voltage driven memristor ($n=-1$), there is only one stationary point $T_{s}$
and no hysteresis is observed. In contrast, for a current driven memristor 
($n=1$), there are either one (stable) or three (two stable and one unstable)
stationary points. The equation $\kappa T_{s}/q_{0}=\bar{r}(T_{s})$ is
graphically solved in Fig. \ref{F2}(a). By simulating Eq. (\ref{1}) at
constant $T$, we determined the curves $\bar{r}(T)$ (red solid curve). For 
$q_{0}^{(1)}<q_{0}<q_{0}^{(2)}$ [$q_{0}^{(1)}$ and $q_{0}^{(2)}$ correspond
to the upper and lower dashed straight lines in Fig. \ref{F2}(a)], a
straight line $\kappa T/q_0$ (dotted) crosses ${\bar{r}(T)}$ in three
points; for all other $q_0$ the straight line and the curve $\bar{r}(T)$
cross in one point, only. The profile of the curve ${\bar{r}}(T)$ can be
sketched analytically by noticing that the average $\langle r(x)\rangle $
must be taken with respect to the Boltzmann distribution density $\rho (x)=e^{-\omega
_{0}^{2}x^{2}/2kT}/\int_{-1}^{1}dxe^{-\omega _{0}^{2}x^{2}/2kT}$ for $|x|<1$
and zero overwise. It follows immediately that the curves ${\bar{r}}(T)$
grow exponentially like $\exp [T/(2\omega _{0}^{2}\lambda ^{2})]$ at low
temperatures, $T/\omega _{0}^{2}\ll 1$, and saturate at $\lambda \sinh
(1/\lambda )$ for high temperatures, $T\gg \omega _{0}^{2}$, in agreement
with our simulations [Fig \ref{F2}(a)]. Therefore, within the interval 
$q_{0}^{(1)}<q_{0}<q_{0}^{(2)}$, one can expect two stationary solutions of
Eqs. (\ref{1}) and (\ref{2}) and hysteresis of both temperature and
resistance, when the applied electrical current is cycled. These conclusions
are confirmed by the numerical integration of the coupled Eqs. (\ref{1}) and
(\ref{2}) for $\kappa =0.01$ [Fig. \ref{F2}(b)]. The characteristics curve
displayed there, $r(q_{0}^{1/2})$ (note that $q_{0}^{1/2}\propto I$),
exhibits hysteretic behavior, thus suggesting that this system can serve as
a memory bit with two logic states corresponding to two different
resistances. However, such a bit is not a non-volatile memory element, yet,
since the two states exist only for a nonzero current (i.e., $q_{0}>0$).

The resistance hysteretic loop gradually shrinks as the position $x_{0}$ of
the potential minimum of $U(x)=\omega _{0}^{2}(x-x_{0})^{2}/2$ shifts away
from the sample center, $x=0$ where $r(x)$ has its minimum. More
interestingly, negative differential resistance (NDR) was observed for 
$x_{0}\gtrsim \lambda $. Indeed, at relatively large $x_{0}$ [Fig. \ref{F1}(c)], 
the resistance first decreases with the current [Fig. \ref{F2}(c), top
red curve] and then increases for higher currents. With decreasing $x_{0}$,
the NDR region shrinks and finally disappears; on further lowering $x_{0}$,
resistance hysteretic loops open up and expand (Fig. 2(c),
 two bottom curves). NDR is
due to the fact that on increasing $T$ the sharp peak of $\rho (x)$ (which
is centered around the potential minimum but far away from the $r(x)$
minimum) broadens; correspondingly, on increasing the current the particle
is allowed to dwell longer in the vicinity of the resistance minimum at $x=0$.

In order to use our device as a \emph{non-volatile memory element}, the
nanoparticle must be confined to a potential $U(x)$ with two wells at least.
Below we analyzed in detail the case of the potential $U=\alpha
x^{2}/2+\beta x^{3}/3+\gamma x^{4}/4$ with two minima. Parameters 
$\alpha,\beta ,\gamma $ control locations and depth of the minima. Hereafter, we
will consider either even potentials with $\beta =0$ [Fig. \ref{F1}(b)] or
asymmetric potentials [Fig. \ref{F1}(c)] with left minimum in the middle, 
$x=0$, and right minimum at $x\approx 0.9$, i.e., near the r.h.s. electrode.

For slow temperature relaxation, $\kappa \ll 1$, Fig. \ref{F2}(d) clearly
shows how the resistance hysteresis gets suppressed with increasing the
depth of the potential minima at $x\approx 0.9$: The looped blue curves of 
${r}$ vs. $q_{0}^{1/2}$ evolve into univalued curves -- a \textquotedblleft
virgin\textquotedblright\ branch showing up only if the nanoparticle was
initially located in the central well. Therefore, current cycling (or
current pulse) allows switching the system from the low (nanoparticle
sitting in the central well) to the high resistive state (particle being
located near an electrode), but \emph{not vice versa!}

\begin{figure}[tbp]
\includegraphics[width=9cm]{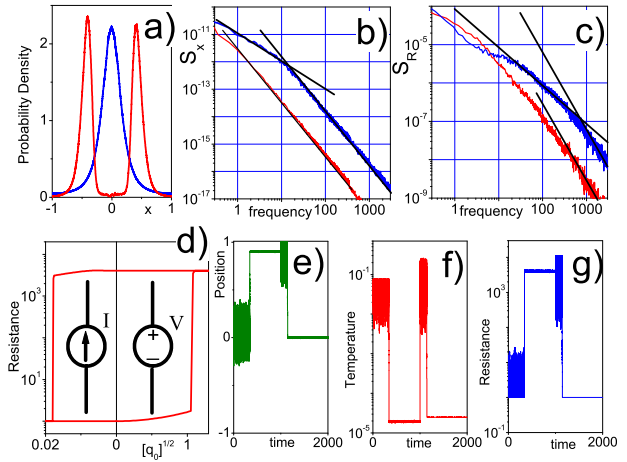}
\caption{Fast temperature relaxation. (a) Spatial probability distribution
of the particle in the current driven switch of Eqs. (\ref{1}), 
(\ref{2}) with double-well potential $U(x)$ with $(\alpha,\beta ,\gamma )=(1,0,4)$. 
The two-peaked (red) and
one-peaked curve (blue) represent the low ($q_{0}^{1/2}=0.1$) and high
current ($q_{0}^{1/2}=0.6$) regime, respectively (see text); (b) Spectra of
the $x(t)$-fluctuations at low and high current as in (a). The power laws 
$1/f$ and $1/f^{2}$ are drawn as appropriate for reader's convenience. (c) As
in (b) but for the $r(t)$ fluctuations. Here, the $1/f$ branch of the
spectrum is more pronounced due to strong nonlinearity of the $r(x)$
function; (d) Resistance cycling between low and high resistive state for
the double-well potential [Fig. \ref{F1}(c)], with $(\alpha, \beta, \gamma )=(4,13.4,8)$. 
The voltage, $[q_{0}]^{1/2}\propto V$, was increased from $0$ to $3$ on the r.h.s. of the
loop and the current, $[q_{0}]^{1/2}\propto I$, from $0$ to $0.02$ on the
l.h.s. Time-dependence of the particle location (e),
temperature (f) and resistance (g) when the voltage pulse, $q_{0}^{1/2}=2.5$, 
applied for $0<t<1000$, is followed by the current pulse, 
$[q_{0}]^{1/2}=0.01$, for $1000<t<2000$ (see text).}
\label{F3}
\end{figure}

\textit{Fast temperature relaxation.---\/} Let us consider now the case of
fast temperature relaxation, $\kappa \gg 1$. For a qualitative analysis, we
can assume that the temperature is always stationary with $%
T(x)=q_{0}[r(x)]^{n}/\kappa $. Substituting this expression into Eq. (\ref{1}%
) for the particle diffusion, we derive the Stratonovich stochastic equation 
\begin{equation}
dx/dt+\partial U/\partial x=[\cosh (x)]^{n/2}[2q_{0}/\kappa ]^{1/2}\xi ,\ \
|x|<1.  \label{4}
\end{equation}%
This equation can be regarded as a model for a Brownian particle diffusing
through a medium with spatially varying temperature (or diffusion
coefficient). For a current driven memristor, $n=1$, the temperature reaches
its maximum when the particle is located near the electrodes, $x=\pm 1$,
while for the a voltage driven memristor the highest temperature occurs when
the particle sits around $x=0$. Therefore, by applying current or voltage
pulses, we can \textquotedblleft heat up \textquotedblright\ the center of
the device or the regions near its electrodes, respectively. This effect can
be invoked to reliably switch the device between low and high resistive
state, with little regard for the parameters of the current or voltage
pulses. Thus, a key problem in memristor design can be solved based on the
present proposal.

We now propose a switching protocol for our model memristor with asymmetric
potential $U$ having minima in the center, $x=0$, and close to the right
electrode, $x\approx 0.9$, namely $\alpha =4$, $\beta =13.4$, and $\gamma =10$. Let
us assume that the system is initially in the low resistive state. With
increasing voltage, $q_{0}^{1/2}\propto V$, the memristor switches to the
high resistive state as shown in our simulations [Fig. \ref{F3}(d); right
side of the loop driven by a voltage source]. Upon decreasing the voltage to
zero, the system sets in its high resistive state, indicating that the
particle is now trapped near the right electrode. If we next increase the
current, $q_{0}^{1/2}\propto I$, the system switches to the low resistive
state [Fig. \ref{F3}(d); left side of the loop driven by current source]. By
finally turning the current off, the resistance remains set to its low
value. The loop being almost deterministic, the proposed switching protocol
is clearly repeatable.

In order to illustrate the underlying switching dynamics, we computed $x(t)$, 
$T(t)$, and $R(t)$ by numerically integrating the coupled Eqs.~(\ref{1})
and (\ref{2}) and switching $n$ from $-1$ (voltage driven) to $1$ (current
driven) at $t=1000$. This mimics the sequence of voltage pulse, with
duration $0<t<1000$, followed by a current pulse for $1000<t<2000$ [Fig. 
\ref{F3}(e)-(f)]. Under the action of a constant voltage, $0<t<1000$, the
initial low resistive state results in strong Joule heating, $V^{2}/R,$ and,
therefore, high temperature [Fig. \ref{F3}(f)] and strong fluctuations of
the nanoparticle position [Fig. \ref{F3}(e)] just after applying a voltage
pulse. However, as soon as the particle hops over the barrier and gets
trapped in the well near the right electrode, the resistance increases and
the heat production drops so fast that the particle temperature and its
spatial fluctuations get quenched [Fig. \ref{F3}(e)-(f) for $t<1000$]. No
further dynamics was observed over time until after the current pulse was
applied for $t>1000$. Therefore, this switching mechanism is robust with
respect to voltage pulse duration: Indeed, while the memristor is driven by
the \emph{voltage pulse}, the particle cannot hop back to the minimum at $x=0$, 
being the temperature felt by the particle near the electrode very low.
When a \emph{current pulse} is applied, $1000<t<2000$, the heat production
starts anew because the particle is now in a high resistive state and
Joule's heating is proportional to the switch resistance, $Q=I^{2}R$. As a
result, the temperature sharply increases and the particle starts to
fluctuate around the minimum near the electrode [Fig. \ref{F3}(e) for $t$
just after $t=1000$]. Eventually, the particle jumps into the central
potential minimum where the resistance is the lowest. At that very moment
the heat production drops and the particle gets stuck at $x=0$; no more
changes are expected as the particle temperature is now very low. A strong
enough current pulse, irrespective of its duration, suffices to switch the
system back to its low resistive state and keep it there.

Pulse driven switching in our model turns out to be so effective because
resistance and heat production strongly correlate with the particle location
and the particle is always attracted toward the region with the lowest heat
(or entropy) production, where it gets trapped. This conclusion can be
easily extended to the case of any memristor with two distinct resistive
states with resistances $R_{1}$ and $R_{2}$ such that $R_{1}\ll R_{2}$. As a
non-equilibrium thermodynamic system tends to attain a state of lowest
entropy production, a voltage pulse will switch the memristor to the high
resistive state $R_{2}$, while a current pulse will switch it back to the
low resistive state $R_{1}$. Thus, our conclusions about optimal switching
sequence are very general and can be readily extended to more elaborated
memristor models.

Finally, we stress that the memristor switching mechanism proposed here can
be regarded as an instance of noise-induced phase transition \cite{noise-induced}. 
Such a stochastic dynamical phenomenon takes place when the
minimum of the temperature $T(x)$ does not coincide with the minimum of the
binding potential, $U(x)$: The Brownian particle tends to dwell about the
potential minima at low temperatures, and about the temperature minima at
high temperatures, thus switching location on increasing the noise. In order
to detect a noise-induced transition in our memristor model, we simulated
the coupled Eqs.~(\ref{1}) and (\ref{2}) for a symmetric double well
potential $U(x)$ with $\alpha =1,\beta =0$, and $\gamma =4$, in the current
driven regime with $\kappa =25$ (fast temperature relaxation) [Fig. \ref{F3}(a)]. 
For weak currents, the stationary temperature $T(x)$ is low and the
particle is mostly localized around the potential minima; the corresponding 
$x$ distribution, $\rho (x)$, is doubly peaked (red curve). At higher
currents, the temperature increases until the particle is pushed out of the
potential wells toward the center of the sample, where the potential has a
maximum, but the temperature is much lower. In the fast relaxation regime of
Eq.~(\ref{4}) such a transition is well described by stationary probability distribution 
$\rho (x)=[\mathcal{N}/\sqrt{T(x)}]\exp[-\int_{0}^{x}dxU^{\prime }(x)/T(x)]$, 
where $U^{\prime }=dU/dx$ and 
$\mathcal{N}$ is a normalization constant. A current increase is accompanied
by a temperature surge, so that the minimum of the distribution at $x=0$ can
change to a maximum \cite{noise-induced}. More interestingly, the spectra of
the resistance and position fluctuations (also called noise) show a
qualitative change in correspondence with such a transition: The standard 
$1/f^{2}$ spectra break up into a low-frequency $1/f$ branch and a
high-frequency $1/f^{2}$ tail [Fig. \ref{F3}(b),(c)]. The unusual
experimental noise spectra reported in Ref. \cite{6} can thus be interpreted
as the signature of a noise-induced transition occurring in TaO$_{x}$
memristors.

\textit{Conclusion.---\/} The resistive switch model discussed here is
expected to ensure the much sought-for properties of controllability and
reliability one needs to design and prototype switchable nanomemristors for
future flash memory devices. On the other hand, based on numerical
simulation, we claim that its response to external current and voltage
signals reproduces well a wide variety of experimental observations reported
in the literature for real memristors. Moreover, the predicted phenomenon of
negative differential resistance can considerably enhance the functionality
of these nanostructures.

We acknowledge many fruitful discussions with R.S. Williams, who strongly
supported this work. S.E.S. also acknowledges The
Leverhulme Trust for partial support of this research; F.M. aknowledges partial support
from the European Commission, grant No. 256959 (NanoPower).


\begin{references}
\bibitem{norm} i.e. we rescaled $\lambda$ by half the distance between the electrodes.
\end{references}
\end{document}